\documentclass[conference]{IEEEtran}

\usepackage{amsfonts}
\usepackage{amsmath}
\usepackage{float}
\usepackage{indentfirst}
\usepackage{amssymb}
\usepackage{epsfig,latexsym}
\usepackage{wrapfig}
\usepackage[justification=normal]{caption}
\usepackage{times}
\usepackage{subfigure}
\usepackage{graphicx}
\usepackage{cite}
\usepackage{subfigure}
\usepackage{bm}
\usepackage{multirow}
\usepackage{algorithm} 
\usepackage{algorithmic} 
\usepackage{multirow}
\usepackage{color}
\usepackage{caption}

\vspace{0cm}  
\ifCLASSINFOpdf
  \else
  \fi

\begin{document}
%
\title{Accurate Energy-Efficient Power Control for Uplink NOMA Systems under Delay Constraint}

\author{\IEEEauthorblockN{Bowen Cai\IEEEauthorrefmark{1}, Yu Chen\IEEEauthorrefmark{1}, Qimei Cui\IEEEauthorrefmark{1}, Xiaoxuan Zhu\IEEEauthorrefmark{2}, Yang Yang\IEEEauthorrefmark{3}}
\IEEEauthorblockA{\IEEEauthorrefmark{1}National Engineering Lab for Mobile Network Technologies}
\IEEEauthorblockA{Beijing University of Posts and Telecommunications, Beijing, China}
\IEEEauthorblockA{\IEEEauthorrefmark{2}China Science and Technology Exchange Center}
\IEEEauthorblockA{\IEEEauthorrefmark{3}Shanghai Research Center for Wireless Communications}
Email: \{cbw3011474, yu.chen, cuiqimei\}@bupt.edu.cn}

\maketitle

\begin{abstract}

Machine-type communications (MTC) devices in 5G will use the Non-orthogonal multiple access (NOMA) technology for massive connections. These devices switch between the transmission mode and the sleep mode for battery saving; and their applications may have diverse quality of service (QoS) requirements. In this paper, we develop a new uplink energy-efficient power control scheme for multiple MTC devices with the above mode transition capability and different QoS requirements. By using the effective bandwidth and the effective capacity models, the system's energy efficiency can be formulated as the ratio of the sum effective capacity to the sum energy consumption. Two new analytical models are used in system's energy efficiency maximization problem: 1) two-mode circuitry model and 2) accurate delay-outage approximation model. Simulation shows our proposed scheme is capable of providing exact delay QoS guarantees for NOMA systems.
\end{abstract}


\section{Introduction}

Non-orthogonal multiple access (NOMA) is a power-domain multiplexing technology
that allows users to transmit signals on the same time-frequency resources [1].
This technology supports scenarios with massive connectivity, therefore, it is a
5G candidate technology for uplink massive machine-type communications (MTC) [2].
MTC devices are usually power limited but have applications with diverse quality of service (QoS) requirements, e.g., public monitoring, real-time localization, industrial
automation. Therefore, energy efficient transmission is a crucial requirement in
most cases. Therefore, it is important to design an energy-efficient power
control scheme in uplink NOMA systems with delay QoS constraint.

Extensive research has been done on power control for NOMA systems by taking
link-layer QoS requirements into consideration. Yang et al. [3] and Cai et al.
[4] proposed a power allocation scheme for NOMA systems under minimum rate
constraints. When modeling a wireless communication system as a queueing system,
the delay QoS requirements can be approximated by effective capacity model [5].
Yu et.al [6] used effective model to analyze the performance of a two-user
downlink NOMA network. Furthermore, Choi [7], Liu et.al [8] and Chen et al. [9]
used the effective capacity model and proposed a cross-layer power control policy
to guarantee a certain delay QoS requirement in NOMA systems. However, in these
work, the transmitter is always in transmission mode, which will over estimate
transmit power consumption in their power control schemes. In 2016, Sinaie et al.
[10] proposed a cross-layer power consumption model by considering a two-mode
circuitry (a circuitry works in the transmission mode or sleep mode). Xu et al.
[11], [12] followed Sinaie's work and proposed a new energy efficiency analytical
model by using a two-mode circuitry. The model is simple but only apply to the
point-to-point wireless communication systems, which is not suitable for NOMA
systems.

The aim of this paper is to design an energy-efficient power control scheme with
delay QoS constraint in uplink NOMA systems. By continuing Xu's work, we develop Xu's energy efficiency analytical model in uplink NOMA systems and formulate
an optimization problem to maximize the energy efficiency of NOMA systems under
delay QoS requirements and peak power constraints. We further use the Dinkelbach
method and develop a new power control scheme that can solve the above
optimization problem.

The rest of this paper is organized as follows. Section II describes the uplink NOMA system model as well as the effective bandwidth and the effective capacity model. In section III, we analyze the effective capacity in NOMA systems. In section IV, we will formulate our power control problem under target delay-outage constraints. In section V, the problem will be solved by Dinkelbach algorithm. Performance of the proposed algorithm is evaluated in Section VI by simulations. Section VII summarizes our work.

\section{System Model}
\subsection{NOMA System Mode}
As shown in Fig. 1, we consider an uplink
NOMA system, in which a NOMA base station serves $K$ MTC users or user equipments (UEs) (in this paper, we use the terms MTC users and UE
interchangeably). A UE works either in a transmission mode if there is data to
transmit or in a sleep mode otherwise. Transitions between two modes can be
modeled as a state machine, which is illustrated in Fig. 2. All the UEs utilize 
the same time and frequency resources but different power domain to transmit
data. The received signal at the base station is given by
\begin{equation}
y =\sum \limits_{k = 1}^K \sqrt {P_k^{tx}} {h_k}{s_k} + n,
\label{eq:1}
\end{equation}
where $h_k\left(k\in1,\cdots,K\right)$ is the Rayleigh fading coefficient with unit
variance between the base station and user $k$, $P_k^{tx}$ denotes the transmit
power for user $k$, $s_k$ denotes the transmit signal for user $k$, and $n$ is the
additive white Gaussian noise (AWGN) with mean $\mu$ and variance $\sigma^2$.

\begin{figure}[t]
\begin{center}
\includegraphics[width=8.6cm]{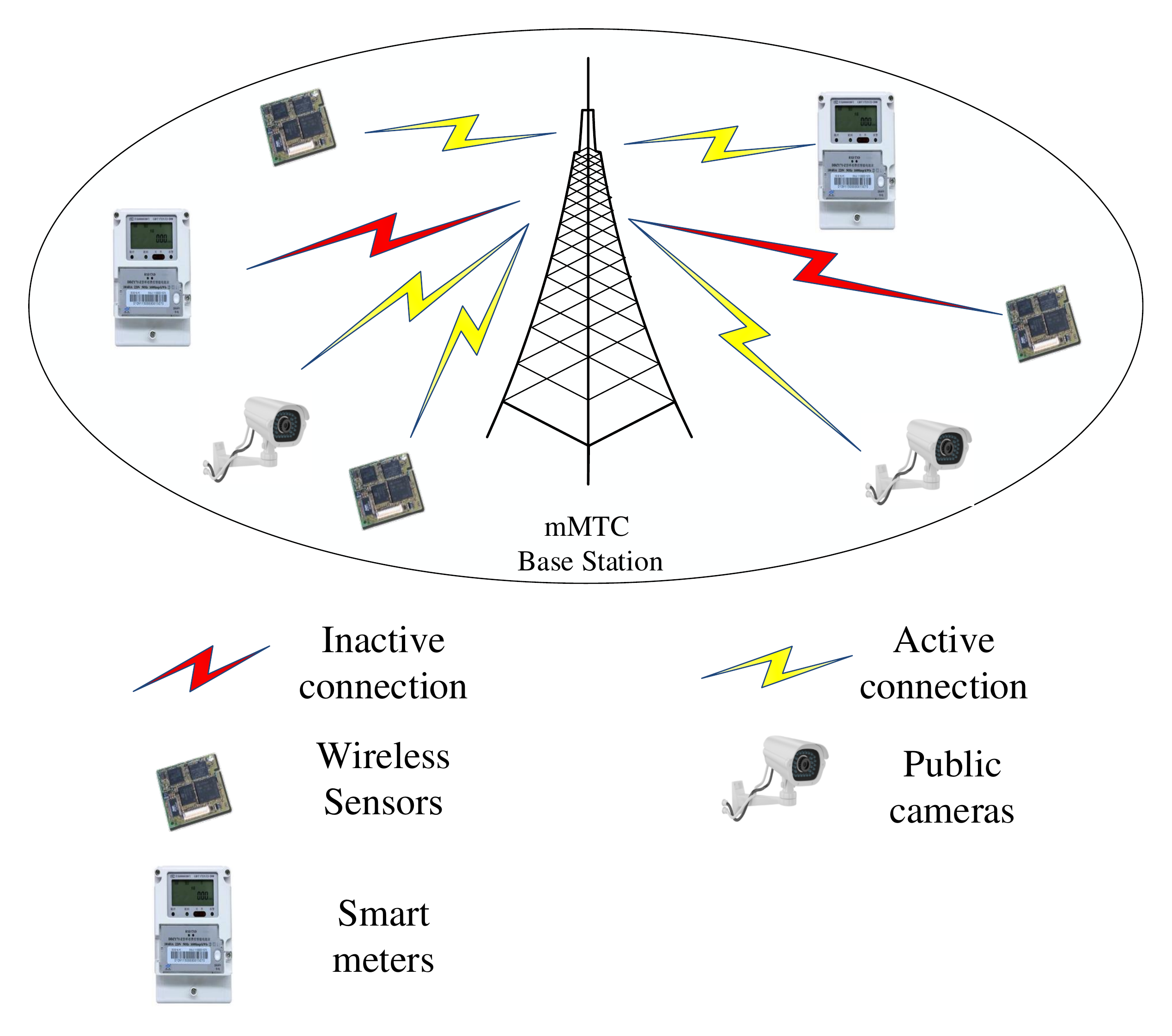}
\caption{System Model}
\label{fig.1}
\end{center}
\end{figure}
\begin{figure}[t]
\begin{center}
\includegraphics[width=8.6cm]{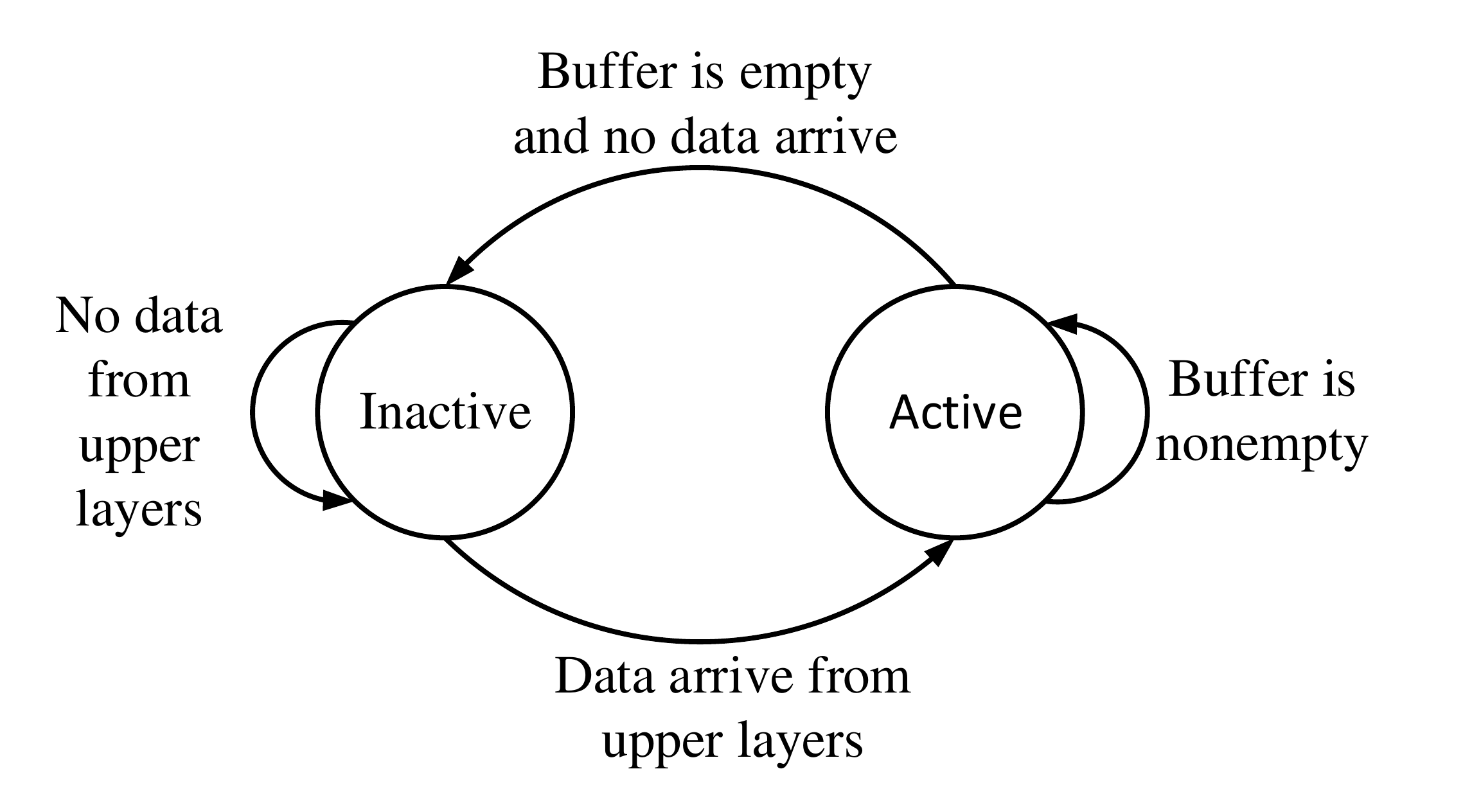}
\caption{Two-mode UE}
\label{fig.2}
\end{center}
\end{figure}

The successive interference cancelation (SIC) technique is used at base station
to eliminate multiuser interference. The SIC decoding order is based on the
channel gain information. The highest channel gain user's signal is decoded at
first since it is strongest at the base station. Then removing the strongest
signal from the received signal to decode the second highest channel gain user's
signal. That is, to decode the $k^{th}$ user's signal first decodes the $i^{th}$
$(i<k)$ user's signal and removes this signal from received signal, in
the order $i=1,2,\cdots,k-1$; the signals from the $i^{th}$ $(i>k)$ users are
treated as noise.

Let ${\left\vert{}h_k\right\vert{}}^2$ denote the channel gain of
the $k^{th}$ user, where $\left|  \cdot  \right|$ is the absolute value of
a complex number. ${\left\vert{}h_k\right\vert{}}^2$ has an independent and
identically distributed exponential distribution with unit mean. For simplicity,
we assume that and the channel gains are sorted in a descending order, i.e., ${\vert{}h_1\vert{}}^2\geq{}{\left\vert{}h_2\right\vert{}}^2\geq{}\cdots{}\geq{}{\vert{}h_K\vert{}}^2$.
Therefore, the signal-to-interference-plus-noise ratio (SINR) experienced when
decoding $k^{th}$ user's signal is
\begin{equation}
\gamma_k = \frac{{{S_k}}}{{{I_k} + {\sigma ^2}}} = \frac{{P_k^{tx}{{\left|
{{h_k}} \right|}^2}}}{{\mathop \sum \limits_{i = k + 1}^K {{\left| {{h_i}}
\right|}^2}P_k^{tx} + {\sigma ^2}}},
\label{eq:2}
\end{equation}
where $S_k$ is the received $k^{th}$ user's signal and
$I_k=\sum_{i=k+1}^K{\left\vert{}h_i\right\vert{}}^2P_i$ represents other users
interference. Specially, the $K^{th}$ user's SINR is
\begin{equation}
{\gamma_K} = \frac{{P_K^{tx}{{\left| {{h_K}} \right|}^2}}}{{{\sigma ^2}}},
\label{eq:3}
\end{equation}

We assume that the instantaneous channel gain information is perfectly known at
each UE. Based on the Shannon theory, the achievable rate of the $k^{th}$ user $C_k$ is
given by
\begin{equation}
{C_k} = B{\log _2}\left( {1 + {\gamma _k}} \right),
\label{eq:4}
\end{equation}
where $B$ is the total bandwidth utilized by all the users.

\begin{figure}[t]
\begin{center}
\includegraphics[width=8.6cm]{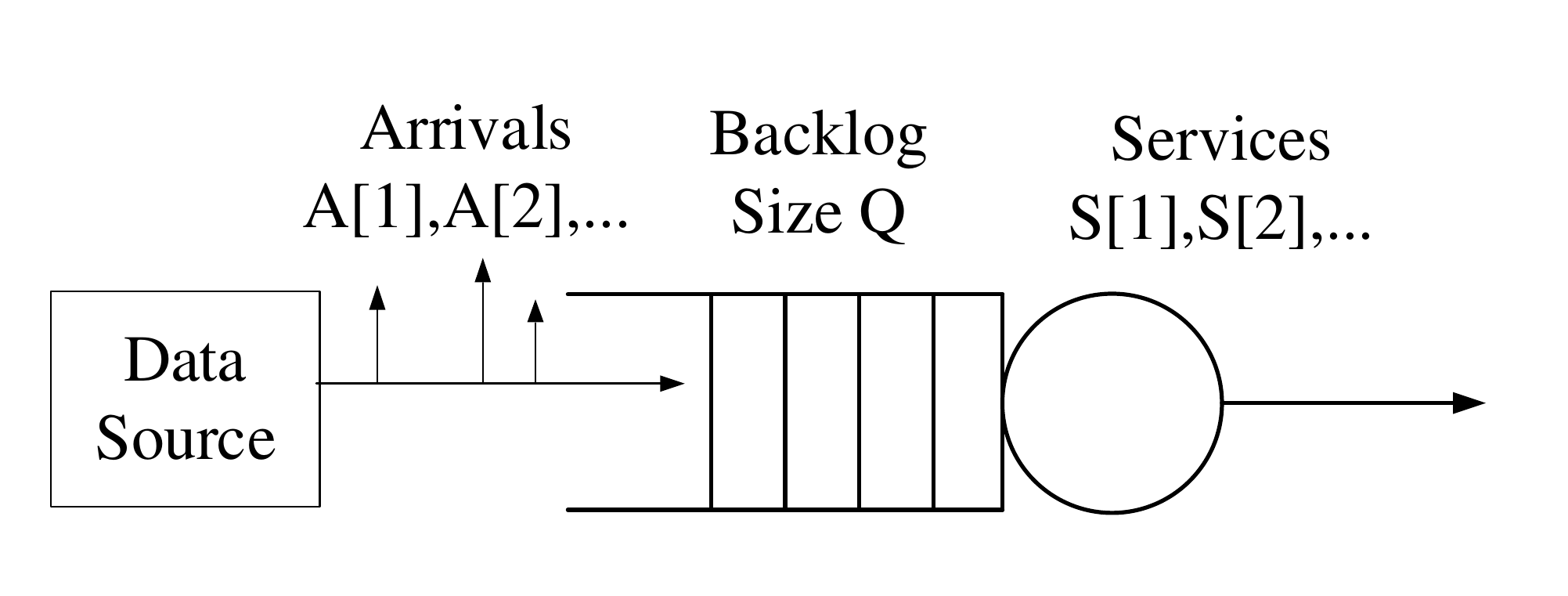}
\caption{Buffer Model}
\label{fig.3}
\end{center}
\end{figure}

\subsection{The preliminaries of effective bandwidth and the effective capacity model}

All the NOMA UEs have its own buffer with infinite buffer size. The buffer model
is shown at Fig. 3. The $k^{th}$ UE arrival data from the data source at slot $n$ is
$A_k\left[n\right]\ (k=\left\{1,2,\cdots,K\right\},n=\{1,2,3,\cdots\})$. Furthermore, we
follow the work [11] and assume that \begin{enumerate}
	\item the $k^{th}$ UE source arrival data confirms to a Bernoulli process with a data
arrival probability $p_k\ (p_k\ \in{}(0,1])$
	\item and the arrival $A_k\left[n\right]$ is exponentially distributed with a mean
data length $L$.
\end{enumerate}

Based on the above assumptions, the $k^{th}$ UE arrivals $A_k\left[1\right],A_k\left[2\right],\cdots$ are  identically distributed (IID) random variables (RVs) identical to a RV $A_k$; the
probability density function (PDF) of the arrival $A_k$ is
\begin{equation}
f_{A_k}(a) = \left\{ {\begin{array}{*{20}{c}}
{p_k}\frac{1}{L}\exp (-\frac{1}{L}a),&a>0\\
1-{p_k},&a=0
\end{array}} \right.
\label{eq:5}
\end{equation}
the average data arrival rate for UE \textit{k }is
\begin{equation}
{\mu _k} = \frac{{{p_k}L}}{{{T_s}}}.
\label{eq:6}
\end{equation}
where $T_s$ denotes the duration of a slot.

Due to the fact that the service rate in any time-varying wireless channels
fluctuates, we assume that the $k^{th}$ user services
$S_k\left[1\right],S_k\left[2\right],\cdots$ are IID RVs identical
to a RV $S_k$
\begin{equation}
{S_k} = {T_s}{C_k} = {T_s}B{\log _2}(1 + {\gamma _k}).
\label{eq:7}
\end{equation}

In order to guarantee a QoS requirement, the effective bandwidth is
defined as the minimum constant service rate and the effective capacity is
defined as maximum constant arrival rate specified by exponent $u$ [5]. Based on
the effective bandwidth model and the effective capacity model, the $k^{th}$ UE's
data arrival rate and service rate can be characterized by their own effective
bandwidth ${\alpha{}}_k^{(b)}(u_k)$ and effective capacity
${\alpha{}}_k^{(c)}(u_k)$:
\begin{equation}
\alpha _k^{(b)}({u_k}) = \frac{{\log E({{\rm{e}}^{{u_k}{A_k}}})}}{{{u_k}{T_s}}}
= \frac{1}{{{u_k}{T_s}}}log\left( {\frac{{{p_k}}}{{1 - {u_k}L}} + 1 - {p_k}}
\right).
\label{eq:8}
\end{equation}
\begin{equation}
\alpha _k^{(c)}({u_k}) =  - \frac{{\log E({{\rm{e}}^{ -
{u_k}{S_k}}})}}{{u_k{T_s}}} =  - \frac{{\log E({{\rm{e}}^{ - {u_k}{T_s}B{{\log
}_2}(1 + {\gamma _k})}})}}{{{u_k}{T_s}}}.
\label{eq:9}
\end{equation}
where $u_k$ is $k^{th}$ UE's QoS exponent. If the assumptions of the Gartner-Ellis
theorem hold, there is a unique QoS exponent $u_k^*>0$ that satisfies
\begin{equation}
\alpha _k^{(b)}{\rm{(}}u_k^*{\rm{) }} = \alpha _k^{(c)}{\rm{(}}u_k^*{\rm{),}}
\label{eq:10}
\end{equation}
then the complementary cumulative distribution function (CCDF) of backlog size
can be approximated by [5]:
\begin{equation}
P({Q_k} > B) \approx p_k^b{e^{ - u_k^*B}},
\label{eq:11}
\end{equation}
where $p_k^b$ is the nonempty buffer probability for $k^{th}$ UE. The parameter $u_k^*$ plays a
critically important role for statistical QoS guarantees, which indicates the
exponential decay rate of the QoS violation probability. A smaller $u_k^*$
corresponds to a slower decay rate, which implies that the $k^{th}$ user can
only provide a looser QoS guarantee, while a larger $u_k^*$ leads to a faster
decay rate, which means that a more stringent QoS requirement can be supported.
When the arrival $A_k$ is exponentially distributed with a mean data length $L$,
the value of the nonempty buffer probability can be calculated from Xu' s work
[11] as
\begin{equation}
p_k^b = 1 - u_k^*L.
\label{eq:12}
\end{equation}

Denote the $k^{th}$ UE circuits power consumption and transmission probability by
$P_k^c$ and $p_k^{tx}$ respectively. The $k^{th}$ UE total power consumption $P_k$
is
\begin{equation}
{P_k} = P_k^c + p_k^{tx}P_k^{tx},
\label{eq:13}
\end{equation}
The probability of a UE being in transmission mode is equivalent to the
probability that traffic arrives from the upper layer or the buffer storage is
non-empty. Denote the event of traffic arrival by $A$ and the event of buffer is
non-empty by $B$. Since two events are mutually independent,
therefore, $p_k^{tx}$ can be expressed as
\begin{equation}
p_k^{tx} = P(A) + P(B) - P(AB) = {p_k} + p_k^b - p_k^b{p_k}.
\label{eq:14}
\end{equation}

\section{Effective capacity analysis in NOMA}
\subsection{Effective capacity analysis for two-user NOMA}
We first consider a two-user NOMA system, where two users utilize the same
time-frequency resources. This scenario has been extensively studied in many
papers [7], [13], which is called multiuser superposition transmission or
paired NOMA. In a two-user NOMA system, the corresponding distribution of SINR for the first user is
given by:

\textit{Case 1: If second user is in sleep mode, the PDF of the SINR is}
\begin{equation}
{f_1}(x) = \frac{{{\sigma ^2}}}{{P_1^{tx}}}{e^{ - \frac{{{\sigma
^2}}}{{P_1^{tx}}}x}}.
\label{eq:15}
\end{equation}

\textit{Case 2: If second user is in transmission mode, the PDF of the SINR is} [14]
\begin{equation}
{f_2}(x) = \left( {\frac{{{\sigma ^2}}}{{P_1^{tx} + P_2^{tx}}} +
\frac{{P_2^{tx}P_1^{tx}}}{{{{(P_1^{tx} + P_2^{tx}x)}^2}}}} \right){e^{ -
\frac{{{\sigma ^2}x}}{{P_1^{tx}}}}}.
\label{eq:16}
\end{equation}
Therefore, the effective capacity for the first user is given by
\begin{equation}
\begin{aligned}
  \alpha_1^{(c)}({u_1}) =  &- \frac{1}{{{u_1}{T_s}}}\log \left[ {E({e^{ - {u_1}{S_1}}})} \right]  \\
   =  &- \frac{1}{{{u_1}{T_s}}}\log \bigg[ p_2^{tx}\int_0^{ + \infty } {{e^{ - {u_1}{T_s}B{{\log }_2}(1 + x)}}} {f_2}(x)dx\\ &+ 
   (1 - p_2^{tx})\int_0^{ + \infty } {{e^{ - {u_1}{T_s}B{{\log }_2}(1 + x)}}} {f_1}(x)dx\bigg].
\label{eq:17}
\end{aligned}
\end{equation}
Since the second user's signal is only interfered by the noise, therefore, the effective capacity for the second user is
\begin{equation}
\begin{aligned}
\alpha _2^{(c)}({u_2}) &=  - \frac{1}{{{u_2}{T_s}}}\log \left[ {E({e^{ -
{u_2}{S_2}}})} \right]\\
&=  - \frac{1}{{{u_2}{T_s}}}\log \left[ {{e^{ - {u_2}{T_s}B{{\log }_2}(1 +
x)}}\frac{{{\sigma ^2}}}{{P_2^{tx}}}{e^{ - \frac{{{\sigma ^2}}}{{P_2^{tx}}}x}}dx}
\right].
\end{aligned}
\label{eq:18}
\end{equation}

\subsection{Effective capacity analysis for K-user NOMA}
The $k^{th}$ user effective capacity in a NOMA systems depends on the distribution of $\gamma_k$, which is difficult to derive the close-form expression when the
number of user is larger than three. Gu et al. [14] derive a simple method to
reduce the computational complexity to calculate the effective capacity of $k^{th}$ user in a full-interference scenario. But in NOMA systems, due to the SIC mechanism, only the first user's signal interfered by other $K-1$ users' signal. The $k^{th}$ user's signal only interfered with $i^{th}$ $(i>k)$ user's signal. Based
on the above facts and integrated with Gu's work,
${\alpha}_k^{\left(c\right)}(u_k)$ can be calculated as 
\begin{equation}
\alpha _k^{(c)}({u_k})\!=\!-\frac{1}{{{u_k}{T_s}}}\log \left\{ {1\!-\!\int_0^1\!e^{-s}\!\prod\limits_{i \in N,i > k}\!{p_i^{tx}\frac{{{\sigma ^2}}}{{{\sigma ^2} +
s{P_i}}}dt} } \right\} 
\end{equation}
where $N$ denotes the set of transmission mode users and
\begin{equation}
s = \frac{{{\sigma ^2}({2^{ - \frac{1}{{{u_k}B{T_s}}}\ln t}} - 1)}}{{P_i^{tx}}}.
\label{eq:19}
\end{equation}

\section{Problem Formulation of the Power control}
In this work, we use bits per Joule to measure the system's energy efficiency,
which is the ratio of the system's sum effective capacity to the total power
consumption:
\begin{equation}
\eta  = \frac{{\mathop \sum \limits_{k = 1}^K \alpha _k^{(c)}({u_k})}}{{\mathop \sum \limits_{k = 1}^K {P_k}}}.
\label{eq:20}
\end{equation}
Now let us consider the delay-QoS requirement. The total delay $D_k\
$experienced by user $k\ $consists of queueing delay $D_k^q$ and transmission
delay that equals $T_s$:
\begin{equation}
{D_k} = D_k^q + {T_s}.
\label{eq:21}
\end{equation}

When the maximum delay bound $D_{max}$ and a tolerance $\varepsilon$ are specified by a typical MTC application, the system is in delay-outage if it cannot guarantee the following inequality:
\begin{equation}
P\left( {{D_k} > {D_{max}}} \right) \le \varepsilon .
\label{eq:22}
\end{equation}
where $P(\cdot)$ denotes the probability of a random event.

Let $\mathbf{P^{tx}}=[P_1^{tx},P_2^{tx},\cdots,P_K^{tx}]$ and $\mathbf{u}$=[$u_1,u_2,\cdots,u_K$] denote the
vector of transmit power and QoS exponent, respectively. The optimal
power control problem can be formulated as \textbf{P1}:
\begin{equation}
{\bf{P1}}{\rm{: }}\max \;\eta ({{\bf{P}}^{{\bf{tx}}}}{\bf{,u}}),
\label{eq:23}
\end{equation}
\begin{equation}
s.t.\;\;P\left( {{D_k} > {D_{max}}} \right) \le \varepsilon .
\label{eq:24}
\end{equation}
\begin{equation}
P_k^{tx} \le {P_{max}}\;.
\label{eq:25}
\end{equation}
where (25) is the delay-outage constraint and (26) is peak power constraint for
each UE.

\section{Power control Strategy}

In this section, we will solve \textbf{P1} to obtain the optimal power control
strategy for maximizing uplink NOMA energy efficiency.

According to [12], the CCDF of queueing delay $D_k^q$ can be approximated by
\begin{equation}
P(D_k^q > t) = {\left( {\frac{{1 - u_k^*L}}{{1 - u_k^*L + {p_k}u_k^*L}}}
\right)^{\frac{t}{{{T_s}}} + 1}}.
\label{eq:26}
\end{equation}
By substituting $t$ into (27) with $D_{max}-T_s$
\begin{equation}
\begin{aligned}
P\left( {{D_k} > {D_{max}}} \right) &= P({D_k^q} > {D_{max}} - {T_s})\\
&= {\left( {\frac{{1 - u_k^*L}}{{1 - u_k^*L + {p_k}u_k^*L}}}
\right)^{\frac{{{D_{max}}}}{{{T_s}}}}}.
\label{eq:27}
\end{aligned}
\end{equation}
The constraint (25) can be re-written as
\begin{equation}
\begin{aligned}
&{\left( {\frac{{1 - {u_k}L}}{{1 - {u_k}L + {p_k}{u_k}L}}}\right)^{\frac{{{D_{max}}}}{{{T_s}}}}} \le \varepsilon 
\\
\Leftrightarrow &{u_k} \ge \frac{{\beta  - 1}}{{({p_k} + \beta  - 1)L}}.
\end{aligned}
\label{eq:28}
\end{equation}
where $\Leftrightarrow{}$ is the equivalent sign and
$\beta{}={\varepsilon{}}^{-\frac{T_s}{D_{max}\ }}$. Result (29) indicates that
constraint (25) give a lower bound of QoS exponent $u_k$. A  large  value  of 
$u_k$ indicates a stringent delay QoS requirement and thus requires more power
consumption. Based on the above observation, we first have the following result:

\textit{Result 1}: When the average data arrival rate is ${\mu{}}_k$, the energy
efficiency of the uplink NOMA system is a decreasing function of QoS exponent
$u_k$.

For a proof of Result 1, see Appendix A. Based on Result 1, the optimal QoS
exponent $u_k\ $is the boundary value
\begin{equation}
{u_k} = \frac{{\beta  - 1}}{{({p_k} + \beta  - 1)L}}.
\label{eq:29}
\end{equation}

By substituting $u_k^*\ $in (30) into (24) in \textbf{P1}, the problem
\textbf{P1} becomes \textbf{P2}:
\begin{equation}
{\bf{P2}}:{\rm}\max \eta ({{\bf{P}}^{{\bf{tx}}}}{\bf{,}}{{\bf{u}}^{\bf{*}}}),
\label{eq:30}
\end{equation}
\begin{equation}
P_k^{tx} \le {P_{max}}.
\label{eq:31}
\end{equation}
where $\mathbf{u^*}=[u_1^*,u_2^*,\cdots,u_K^*]$ is the optimal QoS exponent vector. The problem
\textbf{P2 }is to find such a $\mathbf{P^{tx}}$ that satisfy the constraint (32).  We have
the following result to formally characterize this problem:

\textit{Result 2}: In an uplink NOMA system, for given QoS exponent the vector of
QoS exponents of different UEs, the sum of effective capacity is concave of the
transmission power $P_k^{tx}$.

For a proof of Result 2, see Appendix B. Based on result 2, the numerator of
$\eta$ in (21) is concave. Since the denominator of $\eta$ is an affine
function, the fractional function $\eta$ is qusi-concave [15]. A fractional
quasi-concave problem can be solved by Dinkelbach's algorithm as a sequence of
parameterized concave problems [16]. Let $q_i^*$ be the optimal value of original
problem, $q_i^*$ can be expressed as
\begin{equation}
q_i^* = \mathop {\max }\limits_{\mathbf{P^{tx}}} \left\{ {\frac{{\mathop \sum \limits_{k = 1}^K {\alpha ^{(c)}}(u_k^*)}}{{\mathop \sum \limits_{k = 1}^K {P_k}}}} \right\}.
\label{eq:32}
\end{equation}

Problem \textbf{P2} can be transformed to the
following parametric concave problem:
\begin{equation}
F({q_i}) = \mathop {\max }\limits_{\mathbf{P^{tx}}} \left\{ {\mathop \sum \limits_{k = 1}^K {\alpha ^{(c)}}(u_k^*) - {q_i}\mathop \sum \limits_{k = 1}^K {P_k}} \right\}.
\label{eq:33}
\end{equation}

The maximal value of $q_i^*$ in (34) is a root of the equation
$F\left(q_i^*\right)=0$. The value of $q_i^*$ can be found by
solving the parameterized problem in (34) according to the Dinkelbach method. For a given $q_i^*$ in (34), we solve the problem as \textbf{P3}:
\begin{equation}
{\bf{P3}}:\mathop {\max}\limits_{\mathbf{P^{tx}}} \left\{ {\mathop \sum
\limits_{k = 1}^K {\alpha ^{(c)}}(u_k^*) - q_i^*\mathop \sum \limits_{k = 1}^K
{P_k}} \right\},
\label{eq:34}
\end{equation}
\begin{equation}
P_k^{tx} \le {P_{max}}.
\label{eq:35}
\end{equation}
The problem \textbf{P3} is a concave optimization problem and constraint (36)
satisfies Slater's condition [15], one that can be efficiently solved by the
Lagrange dual method. The Lagrangian function of problem \textbf{P3} can be
written as
\begin{equation}
L(P_k^{tx},{\lambda _k}) = \mathop \sum \limits_{k = 1}^K {\alpha ^{(c)}}(u_k^*)
- q_i^*\mathop \sum \limits_{k = 1}^K {P_k} + \mathop \sum \limits_{k = 1}^K
{\lambda _k}(P_k^{tx} - {P_{max}}),
\label{eq:36}
\end{equation}
where ${\lambda{}}_k$ is the nonnegative Lagrange multipliers. The equivalent
dual problem can be decomposed into two parts: 1) the maximization problem solves
the power control problem and 2) the minimization problem solves corresponding
Lagrange multiplier, which is given by
\begin{equation}
\mathop {\min }\limits_{{\lambda _k} \ge 0} \mathop {\max }\limits_{P_k^{tx}}
\;L(P_k^{tx},{\lambda _k})
\label{eq:37}
\end{equation}
By using the Lagrange dual decomposition,  the maximization problem can be
solved by differentiating $L(P_k^{tx},{\lambda{}}_k)$with $P_k^{tx}$. We denote
the optimal power control by $P_k^{tx*}$. The Karush--Kuhn--Tucker (KKT)
conditions for \textbf{P3} are given by
\begin{equation}
P_k^{tx*} \le {P_{max}},
\label{eq:38}
\end{equation}
\begin{equation}
\lambda _k^* \ge 0,
\label{eq:39}
\end{equation}
\begin{equation}
\lambda _k^*(P_k^{tx*} - {P_{max}}) = 0,
\label{eq:40}
\end{equation}
\begin{equation}
\frac{{dL}}{{dP_k^{tx*}}} = \frac{B}{{\log 2}}E\left( {\frac{{{\gamma _k}{e^{ -
u_k^*{R_k}}}}}{{P_k^{tx*}(1 + {\gamma _k})E({e^{ - u_k^*{R_k}}})}}} \right) -
q_i^*{p_k} + \lambda _k^* = 0.
\label{eq:41}
\end{equation}
where  $(\cdot)^*$ represents the value of corresponding
variable at the optimal point. Equation (39) and (40) are feasibility conditions,
(41) is the complementary slackness condition, and (42) is the stationary
condition. The optimal power $P_k^{tx*}$ can be solved as
\begin{equation}
P_k^{tx*} = {\bigg[\frac{{B{\gamma _k}}}{{\log 2(1 + {\gamma _k})(q_i^*{p_k} - \lambda
_k^*)}}\bigg]^ + },
\label{eq:42}
\end{equation}
where ${\left[x\right]}^+$ denotes $max⁡\{x,0\}$ and ${\lambda{}}_k^*$ is the
optimal Lagrangian multiplier, which need to ensure the system to meet the peak
power constraint of each UE.

\begin{figure}[t]
\begin{center}
\includegraphics[width=8.6cm]{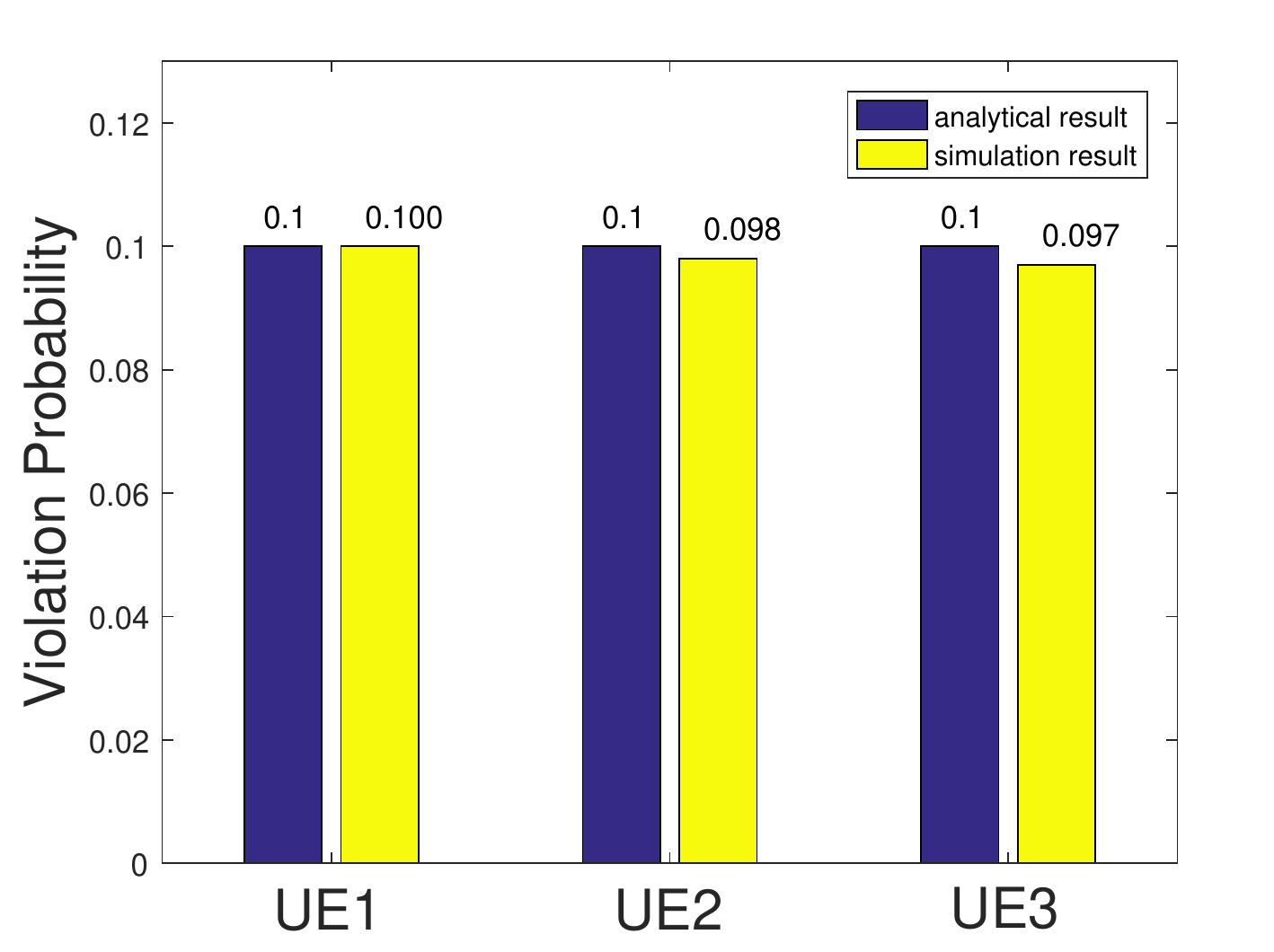}
\caption{Simulation and approximation results of delay violation probability for different UEs.}
\label{fig.4}
\end{center}
\end{figure}

As for minimization problem, the multiplier ${\lambda{}}_k^*$ can be updated by
the subgradient method [15] as follows:
\begin{equation}
{\lambda _k}(j + 1) = {[{\lambda _k}(j) + {\beta _k}(j)({P_{max}} - P_k^{tx*})]^
+ },
\label{eq:43}
\end{equation}
where $j$ is the iteration index, ${\beta{}}_k(j)$ is the positive step sizes for
the $j^{th}$ iteration (e.g., $\frac{1}{\sqrt{j}}$). When the step sizes are chosen
properly, the convergence to the optimal solution is guaranteed.

\begin{figure}[t]
\begin{center}
\includegraphics[width=8.6cm]{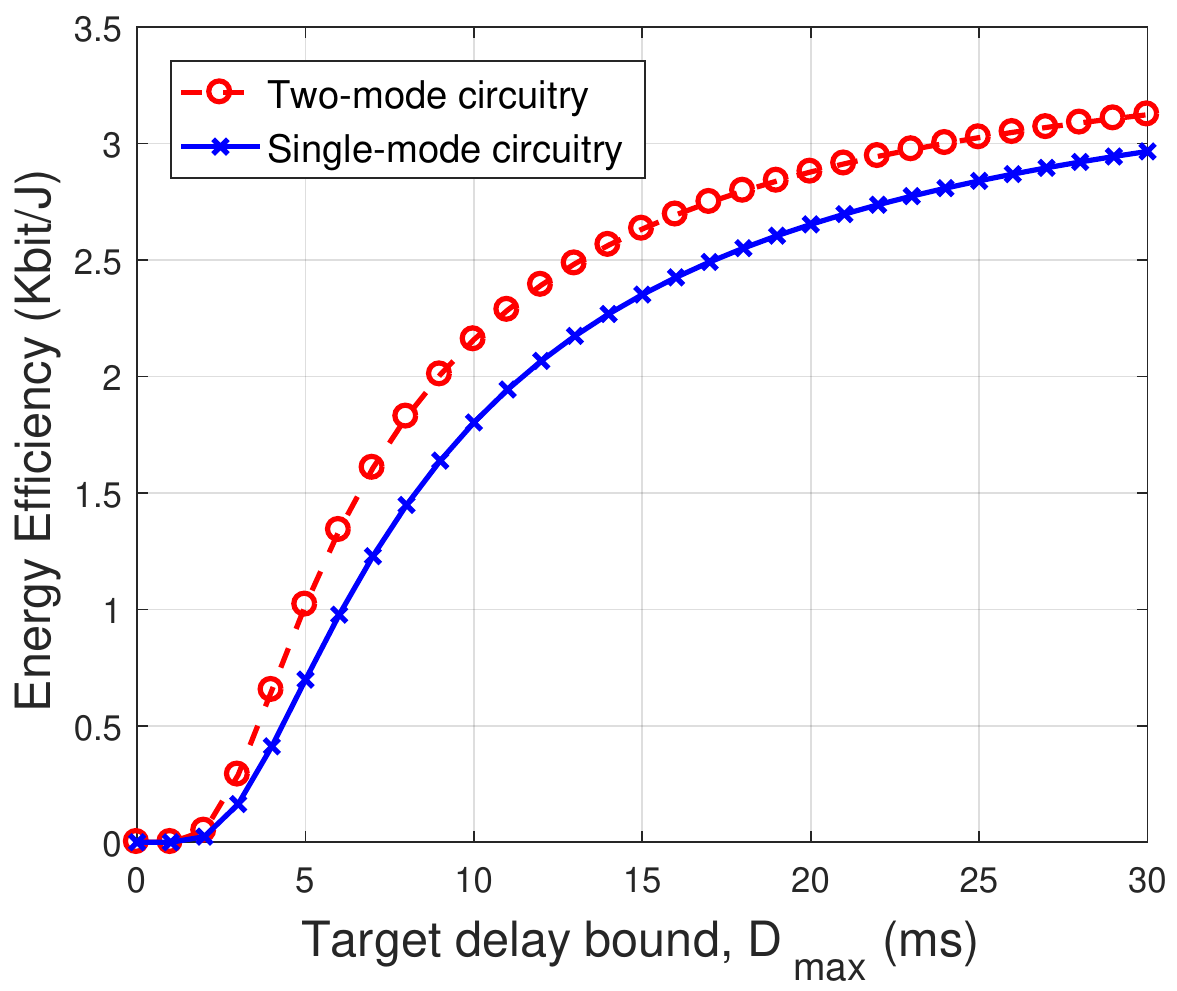}
\caption{Optimal energy efﬁciency under different target delay bound}
\label{fig.5}
\end{center}
\end{figure}

\begin{table}[H]
\setlength{\abovecaptionskip}{0cm}
\setlength{\belowcaptionskip}{0cm}
\centering
\caption{System Parameters}
\renewcommand\arraystretch{1.17}
\begin{tabular}{|l|l|}
\hline
\textbf{Parameters} & \textbf{Values} \\
\hline
Time duration of a slot, $T_s$ & 1ms  \\
\hline
Noise spectral density, $N_0$ & -174dBm/Hz \\
\hline
System Bandwidth, $B_c$ & 18kHz \\
\hline
Circuit power, ${P_{tr}}$ & 10dBm \\
\hline
Maximum transmit power, ${P_c}$ & 46dBm \\ 
\hline
Predefined threshold, $\varepsilon$ & 0.1 \\
\hline
\end{tabular}
\end{table}

\section{Result and discussion}
In this section, we evaluate the performance of the proposed algorithm in
Rayleigh fading environment [17]. The channel gain
${\left\vert{}h_k\right\vert{}}^2$ is generated by exponential distribution with
parameter ${\chi{}}_k=d_k^{\beta{}}$ [18], where $d_k$ is the distance between the
$k^{th}$ user and the base station, $\beta$ denotes the path loss exponent and we
set $\beta{}=4$. Three different QoS requirements UEs are in a NOMA cell. The
distance between base station and three users are 300, 600 and 900 meters,
respectively. The other parameters are listed in Table I.

Fig. 4 shows the simulated and approximated of average delay violation
probability for different UEs when the all the UEs average date rates are 600
Kbps. The delay QoS requirements for three UEs are set as (10ms, 0.1), (20ms,
0.1) and (30ms, 0.1) respectively. As shown in Fig. 4, the approximation results
are very close to the simulation results. This verifies the correctness of
delay-outage probability and indicates that our power control scheme is capable
of providing precise delay-outage probability guarantees.

Fig. 5 shows a comparison of energy efficiency between the two-mode circuitry and
single-mode circuitry. The target delay violation probability $\varepsilon$ is 0.1.
The figure shows the energy efficiency improvement when the two-mode
circuitry is used in NOMA systems. The reason is that the single-mode circuitry
overestimates transmit power, thus it is apparently less energy efficient.

\section{Conclusion}
It is important to design an energy efficient power control scheme in uplink NOMA systems under predefined delay-outage constraints for MTC. Previous work on power control in NOMA systems only consider single-mode UEs, which will over estimate transmit power consumption. In this paper, we consider two-mode UEs in uplink NOMA systems, and propose an energy-efficient power control scheme under predefined delay-outage constraints. Simulation results confirm our power control scheme is capable of providing precise delay-outage probability guarantees in uplink NOMA systems.

\section*{Acknowledgement}
The work was supported in part by the National Nature Science Foundation of China Project under Grant 61471058, in part by the Hong Kong, Macao and Taiwan Science and Technology Cooperation Projects under Grant 2016YFE0122900, in part by the Beijing Science and Technology Commission Foundation under Grant 201702005 and in part by the 111 Project of China under Grant B16006.

\section*{Appendix A}
Since effective capacity ${\alpha}_k^{\left(c\right)}\left(u_k\right)$ is a
decreasing function of $u_k$ [5], therefore, the sum of effective capacity
$\sum_{k=1}^K{\alpha}_k^{\left(c\right)}(u_k)$ is also a decreasing function of
$u_k$. When the QoS exponent $u_k$ increases,  the transmission power $P_k^{tx}$
increases accordingly to meet a stringent QoS requirement. Thus the total power
consumption $P_k$ is an increasing function of the QoS exponent $u_k$. Because
the numerator is a decreasing function and the denominator is an increasing
function, therefore the system's energy efficiency $\eta$ is a decreasing
function of the QoS exponent $u_k$.

\section*{Appendix B}
Let $h\left(x\right)=-\frac{1}{u_kT_s}logE\left(e^{-u_kT_sBlog_2\left(1+c_kx\right)}\right)\left(x>0\right)$, where $c_k=\frac{{\left\vert{}h_k\right\vert{}}^2}{I_k+{\sigma{}}^2}$. Since
${log}_2{\left(1+c_kx\right)}$ is concave for all $x>0$, so $–u_kT_sBlog_2\left(1+c_kx\right)$ is concave in the domain set. This implies that
$e^{-u_kT_sBlog_2\left(1+c_kx\right)}$ is log-convex, and $E\left(e^{-u_kT_sBlog_2\left(1+c_kx\right)}\right)$is log-convex as well since log-convexity is
preserved under sums [15]. Noting that $\log{\left(g\left(\cdot\right)\right)}$ is convex for log-convex
$g\left(\cdot\right)$. Therefore, $h\left(x\right)$ is a concave function of
$x$ for $x>0$. Meanwhile, the sum effective capacity can be written as
$\sum_{k=1}^Kh\left(P_k^{tx}\right)$. Since concavity is preserved under sums,
thus $\sum_{k=1}^Kh\left(P_k^{tx}\right)$ is a concave function of transmission
power $P_k^{tx}$, completing the proof.

\end{document}